\documentclass[conference]{IEEEtran}
\ifCLASSINFOpdf
\else
\fi
\usepackage{amsfonts}
\usepackage[dvips]{graphicx}
\usepackage{times}
\usepackage{cite}
\usepackage{amsmath}
\usepackage{array}
\usepackage{amssymb}

\newcommand{\bs}{\boldsymbol}

\usepackage{stfloats}
\usepackage{slashbox}
\usepackage{graphicx}
\usepackage{footnote}
\usepackage{booktabs}
\usepackage{array}
\usepackage{algorithmic}
\usepackage{algorithm}
\usepackage{subeqnarray}
\usepackage{cases}
\usepackage{threeparttable}
\usepackage{color}
\IEEEoverridecommandlockouts

\begin{document}
%
\title{Joint Secure Beamforming for Cognitive Radio Networks with Untrusted Secondary Users}

\author{Meng Zhang and Yuan Liu
\vspace*{0.5em}\\
School of Electrical and Information Engineering, South China University of Technology, Guangzhou, P. R. China
 \\Email: jackey\_zm@hotmail.com, eeyliu@scut.edu.cn.

\thanks
{This work is supported in part by the Natural Science Foundation of China under Grant 61401159.}}


%


\maketitle

\begin{abstract}
In this paper, we consider a cognitive radio network (CRN) consisting of a primary transmitter-receiver pair and an untrusted secondary transmitter-receiver pair, and each pair is a multiple-input single-output (MISO) link. We consider two transmission schemes, namely underlay scheme and cooperative scheme. For the underlay scheme, the secondary user (SU) is allowed to transmit simultaneously in the presence of primary transmission. For the cooperative scheme, the secondary transmitter acts as a relay node to increase the secrecy rate of primary transmission in exchange for its own transmission. For both schemes, the SU is untrusted and considered as a potential eavesdropper. Our goal is to minimize the total power consumption while satisfying the primary user (PU)'s required secrecy rate and the SU's required information rate. By suitable optimization tools, we design the joint secure beamforming for both schemes. The simulation results show that  in the considered system model, the underlay scheme outperforms the cooperative scheme, especially with high rate requirements and a large number of antennas at secondary transmitter.
\end{abstract}


%
\IEEEpeerreviewmaketitle

\section{Introduction}
Cognitive radio \cite{CRN} was proposed as an efficient method to improve the spectrum efficiency of wireless networks. It allows primary user (PU) networks to share their spectrum with the secondary users (SUs), provided that the SU's transmission does not adversely affect the PU's performance. Usually there are three models of cognitive radio networks (CRN): interweave, underlay and overlay models. In the interweave model, the SUs first sense the spectrum holes and then transmit when the PUs are absent\cite{CR}. In the underlay model, the SUs simultaneously transmit with the PUs over the same spectrum, while maintaining the performance of primary transmission under an acceptable threshold \cite{CR2}. The overlay model enables users cooperation where the SUs aid the PUs' transmission in exchange for their own transmission, thus not only enhancing the spatial reusability but also enlarging the coverage range\cite{CCRN,MISOCR}.
Note that the three models have their own advantages and disadvantages, and are applicable for different scenarios and services.

    \begin{figure}[t]
\begin{centering}
\includegraphics[scale=0.55]{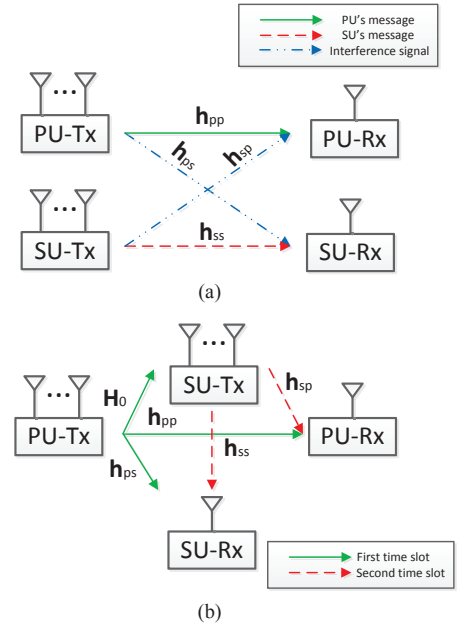}
\vspace{-0.1cm}
 \caption{(a) The underlay scheme and (b) the cooperative scheme for a CRN with an untrusted SU.}\label{fig:sce1}
\end{centering}
\vspace{-0.3cm}
\end{figure}
Nevertheless, PU-SU cooperation based overlay model in CRNs may result in severer security problems than the interweave and underlay models since SUs have to decode the PUs' messages before relaying.
Note that even without the security consideration, which one of the underlay and cooperative-based overlay models is better is case-dependent.
 Then a question arises: if the SUs are untrusted, how are the performance of the underlay and cooperative models?

 To address the security problem, physical-layer security is a promising secure communication means and becomes an emerging area recently \cite{Untrusted,Two-way,SWIPTOFDMA,MengTIFS}. In \cite{Untrusted}, the authors studied the problem of maximizing the secrecy capacity in multiple-input multiple-output (MIMO) one-way relaying systems by joint secure beamforming design at the source and the untrusted relay. It is shown that the secrecy rate can be improved even with the aid of an untrusted relay. The similar problem was studied for MIMO two-way relaying system in \cite{Two-way}. The authors in \cite{SWIPTOFDMA,MengTIFS} studied physical-layer security in OFDMA systems with simultaneous wireless information and power transfer.

 In the CRN scenario, physical-layer security has been studied by avoiding information leakage to a third party (or an external) eavesdropper \cite{SecCog1,SecCog2}. There is another security issue where SUs try to eavesdrop PUs' message without permission, since SUs may easily know PUs' transmission spectrum by spectrum sensing. To our best knowledge, only \cite{UntrustedSU} considered such case, while the authors focused on the achievable rate characterization for cooperative model in single-input single-output (SISO) channel.

In this paper, we study the secure beamforming design for a CRN consisting of a PU pair and an SU pair. Each dedicated transmitter-receiver pair is a multiple-input single-output (MISO) link. The SU is considered to be \textit{untrusted} in the sense it may eavesdrop PU's transmission. We study two transmission schemes, namely \textit{underlay scheme} and \textit{cooperative scheme}.
Our goal is to minimize the total power consumption while satisfying the PU's required secrecy rate and the SU's required information rate. Using suitable optimization tools, we design the joint secure beamforming for both schemes. The simulation results show that the underlay scheme is superior to the cooperative scheme, especially with high required rates of both PU and SU and a large number of SU-Tx's antennas.

\textit{Notations}: Bold upper and lower case letters denote matrices and vectors, respectively. Let $(\cdot)^H$ denote the conjugate transpose.  The matrix $\textbf{I}_N$ is an $N \times N$ identity matrix and the matrix $\textbf{0}$ is an all-zero matrix with appropriate dimensions. For matrix $\bs X$, vec$(\bs X)$, tr$(\bs X)$ represent the vectorization and trace of matrix $\bs X$, respectively. $\bs X\otimes \bs Y$ stands for the kronecker product of $\bs X$ and $\bs Y$. For a vector $\bs x$, we use $||\bs x||_2$ to denote its two-norm.
\section{System Model and Problem Formulation}

We consider a CRN consisting of four nodes: a primary transmitter (PU-Tx), a primary receiver (PU-Rx), a secondary transmitter (SU-Tx) and a secondary receiver (SU-Px) as shown in Fig. 1. PU-Tx and SU-Tx are equipped with $N_p\geq2$ and $N_s\geq2$ antennas, respectively; while each of the receivers is equipped with single antenna. The SU is untrusted in the sense it may disguise as an innocent user and attempt to decode PU's information. Thus, the SU is considered to be a potential eavesdropper.


Let $\mathbf{H}_0 \in \mathbb{C}^{N_{S}\times N_{P}}$ denote the channel of PU-Tx to SU-Tx link, $\mathbf{h}_{pp}\in \mathbb{C}^{N_{P}\times 1}$ the channel of PU-Tx to PU-Rx link, $\mathbf{h}_{sp}\in \mathbb{C}^{N_{S}\times 1}$ the channel of SU-Tx to PU-Rx link and $\mathbf{h}_{ps}\in \mathbb{C}^{N_{p}\times 1}$ the channel of PU-Tx to SU-Rx link. we assume full channel state information (CSI) are available at all nodes. We also assume that all nodes operate in half-duplex mode for the practical consideration.


In the paper, we study two schemes described in the following.

\subsection{Underlay Scheme}
First, we consider an underlay scheme as shown in Fig. 1(a), where the PU-Tx and the SU-Tx simultaneously transmit information to their dedicated receivers in a spectrum-sharing manner. The signals received at the PU-Rx and SU-Rx are respectively given by
\begin{align}
y_p^u&=\mathbf{h}_{pp}^H\mathbf{w}_1{s}_1+\mathbf{h}_{sp}^H\mathbf{w}_2{s}_2+n_p,\\
y_s^u&=\mathbf{h}_{ss}^H\mathbf{w}_2{s}_2+\mathbf{h}_{ps}^H\mathbf{w}_1{s}_1+n_s,
\end{align}
where ${s}_1$ and ${s}_2$ are the information-carrying symbol of PU-Tx and SU-Tx, respectively; $\mathbf{w}_1$ and $\mathbf{w}_2$ represent the transmit beamforming vectors of PU-Tx and SU-Tx, respectively; and $n_p$ and $n_s$ denotes the the additive white Gaussian noise  (AWGN) following $\mathcal{CN}(0,\sigma^2)$ at PU-Rx and SU-Rx, respectively.

For the underlay scheme, the SU-Rx is capable of eavesdropping PU-Tx's transmission and the secrecy rate of PU thus is given by \cite{InterSec}
\begin{align}
R_p^{s,u}=&\left[\log_2\left(1+\frac{\mathbf{w}_1^H\mathbf{h}_{pp}\mathbf{h}_{pp}^H\mathbf{w}_1}{\sigma^2+\mathbf{w}_2^H\mathbf{h}_{sp}\mathbf{h}_{sp}^H\mathbf{w}_2}\right)\right.\nonumber\\
&\left.-\log_2\left(1+\frac{\mathbf{w}_1^H\mathbf{h}_{ps}^H\mathbf{h}_{ps}\mathbf{w}_1}{\sigma^2}\right)\right]^+,
\end{align}
where $[\cdot]^+\triangleq\max(0,\cdot)$. The information rate of SU is given by
\begin{align}
R_s^u=&\log_2\left(1+\frac{\mathbf{w}_2^H\mathbf{h}_{ss}\mathbf{h}_{ss}^H\mathbf{w}_2}{\sigma^2+\mathbf{w}_1^H\mathbf{h}_{ps}\mathbf{h}_{ps}^H\mathbf{w}_1}\right).
\end{align}

Our goal is to design the PU's and SU's beamforming $\mathbf{w}_1$ and $\mathbf{w}_2$ to minimize the total power consumption and the formulated problem is given by
\begin{align}
{\rm (P1)}~\min_{\mathbf{w}_1,\mathbf{w}_2} &\mathbf{w}_1^H\mathbf{w}_1+\mathbf{w}_2^H\mathbf{w}_2\nonumber\\
{\rm s.t.}~~ &R_p^{s,u}\geq Q_p\label{con1:secrecyrate}\\
     &R_s^u\geq Q_s\label{con1:rate}
\end{align}
where $Q_p$ is the minimal secrecy rate requirement for PU and $Q_s$ is the minimal information rate requirement for SU.
\subsection{Cooperative Scheme}
For cooperative scheme, the SU is willing to help the PU transmission using amplify-and-forward (AF) relaying to access the channel but considered to be untrusted.

The cooperative scheme consists of two time slots, as shown in Fig. \ref{fig:sce1}(b). In the first time slot, the PU-Tx transmits its signal to the SU-Tx which is also received by the PU-Rx and SU-Rx. The received signals at SU-Tx, SU-Rx and PU-Rx in the first time slot are respectively given by
\begin{align}
\mathbf{y}_r&=\mathbf{H}_0\mathbf{w}_1{s}_1+\mathbf{n}_r,\\
y_s^c&=\mathbf{h}_{ps}^H\mathbf{w}_1{s}_1+n_s,\\
y_{p,1}^c&=\mathbf{h}_{pp}^H\mathbf{w}_1{s}_1+n_p,
\end{align}
where $\mathbf{n}_r$ represents the AWGN vector at SU-Tx following the distribution $\mathcal{CN}(\mathbf{0},\sigma^2\mathbf{I}_{N_s})$.

In the second time slot, the SU-Tx amplifies and forwards the PU-Tx's signal, and simultaneously transmits its own signal to SU-Rx while PU-Tx remains silent. The received signals at PU-Rx and SU-Rx in the second time slot are respectively given by
\begin{align}
y_{p,2}^c&=\mathbf{h}_{sp}^H\mathbf{F}\mathbf{y}_r+\mathbf{h}_{sp}^H\mathbf{w}_2{s}_2+n_p\nonumber\\
&=\mathbf{h}_{sp}^H\mathbf{F}(\mathbf{H}_0\mathbf{w}_1{s}_1+\mathbf{n}_r)+\mathbf{h}_{sp}^H\mathbf{w}_2{s}_2+n_p,\\
y_s^c&=\mathbf{h}_{ss}^H\mathbf{w}_2{s}_2+\mathbf{h}_{ss}^H\mathbf{F}\mathbf{H}_0\mathbf{w}_1{s}_1+\mathbf{h}_{ss}^H\mathbf{F}\mathbf{n}_r+n_s.\label{ysc}
\end{align}

Note that if SU-Rx can successfully receive message sent from PU-Tx in the first time slot, then the interference term $\mathbf{h}_{ss}^H\mathbf{F}\mathbf{H}_0\mathbf{w}_1{s}_1$ can be subtracted from \eqref{ysc} using the received signal to improve secondary transmission.

In this scheme, the PU-Rx receives two independent copies of the signal transmitted by the PU-Tx in two time slots, respectively. The first copy of the signal is from the direct transmission by the PU-Tx in the first time slot and the second is forwarded by the SU-Tx in the second time slot. By maximal ratio combining (MRC) to these two signals, the PU-Rx can thus retrieve the PU-Tx's information.

Here we consider the worst-case scenario where the SU-Tx and SU-Rx are perfectly colluding, i.e., the output of the wiretap channel is the collection of signals received by the SU-Tx and SU-Rx. Thus, the PU-Tx to the colluding eavesdroppers link and signal-to-noise ratio (SNR) of the colluding wiretap channel are denoted by
\begin{equation}
\mathbf{H}_{e}=\left(
\begin{array}{c}
\mathbf{H}_0\\
\mathbf{h}_{ps}^H
\end{array}\right),~~ \gamma_e=\frac{\mathbf{w}_1^H\mathbf{H}_e^H\mathbf{H}_e\mathbf{w}_1}{\sigma^2}.
\end{equation}

Thus, the secrecy rate for PU and the information rate for SU are respectively given by
\begin{align}
R_p^{s,c}&=\frac{1}{2}\left[\log_2\left(1+\gamma_p\right)-\log_2(1+\gamma_e)\right]^+,\\
R_s^c&=\frac{1}{2}\log_2(1+\gamma_s),
\end{align}
where factor $\frac{1}{2}$ results from the half-duplex transmission mode; $\gamma_p$ and $\gamma_s$ are the signal-to-interference-and-noise ratio (SINR) at PU-Rx and SU-Rx, respectively given by
\begin{align}
\gamma_p=&\frac{\mathbf{h}_{pp}^H\mathbf{w}_1\mathbf{w}_1^H\mathbf{h}_{pp}}{\sigma^2}\nonumber\\
&+\frac{\mathbf{h}_{sp}^H\mathbf{F}\mathbf{H}_0\mathbf{w}_1\mathbf{w}_1^H\mathbf{H}_0^H\mathbf{F}^H\mathbf{h}_{sp}}{\mathbf{h}_{sp}^H\mathbf{w}_{2}\mathbf{w}_{2}^H\mathbf{h}_{sp}+\sigma^2(1+\mathbf{h}_{sp}^H\mathbf{F}\mathbf{F}^H\mathbf{h}_{sp})},\\
\gamma_s=&\frac{\mathbf{h}_{ss}^H\mathbf{w}_2\mathbf{w}_2^H\mathbf{h}_{ss}}{\sigma^2(1+\mathbf{h}_{ss}^H\mathbf{F}\mathbf{F}^H\mathbf{h}_{ss})+a\mathbf{h}_{ss}^H\mathbf{F}\mathbf{H}_0\mathbf{w}_1\mathbf{w}_1^H\mathbf{H}_0^H\mathbf{F}^H\mathbf{h}_{ss}},
\end{align}
where $a$ is a binary variable with $a=0$ indicating that SU-Rx correctly receives the signal from PU-Tx in the first time slot and thus can subtract the interference in the second time slot and otherwise $a=1$. In this paper, we assume $a=1$, which is the worst case.

The power consumption of transmit power for PU-Tx, SU-Tx to relay PU-Tx's information, and SU-Tx to transmit its own information can be obtained as
\begin{align}
P_p=&\frac{1}{2}\mathbf{w}_1^H\mathbf{w}_1,\\
P_r=&\frac{1}{2}{\rm tr}\{\mathbf{F}\mathbf{H}_0\mathbf{w}_1\mathbf{w}_1^H \mathbf{H}_0^H\mathbf{F}^H+\sigma^2\mathbf{F}\mathbf{F}^H\},\\
P_s=&\frac{1}{2}\mathbf{w}_2^H\mathbf{w}_2.
\end{align}

Thus, with the aim to jointly design $\mathbf{w}_1$, $\mathbf{w}_2$ and $\mathbf{F}$, the power minimization problem for the cooperative scheme is formulated as
\begin{align}
{\rm (P2)}~\min_{\mathbf{w}_1,\mathbf{w}_2,\mathbf{F}} &P_p+P_r+P_s\nonumber\\
{\rm s.t.}~~ &R_p^{s,c}\geq Q_p\label{con:secrecyrate1}\\
     &R_s^c\geq Q_s\label{con:rate}.
\end{align}

\section{Beamforming Designs for Underlay Scheme}

To solve (P1) optimally, we define $\mathbf{w}\triangleq[\mathbf{w}_1^H ~\mathbf{w}_2^H]^H$ so as to jointly optimize $\mathbf{w}_1$ and $\mathbf{w}_2$. For this problem, it can be shown that there always exists an SINR constraint $\beta_1$ at PU-Rx such that the following quadratically constrained quadratic problem (QCQP)
\begin{align}
{\rm (P1.1)}~~~\min_{\mathbf{w}} ~~&\mathbf{w}^H\mathbf{w}\nonumber\\
{\rm s.t.}~~~ &\mathbf{w}^H\mathbf{B}_1\mathbf{w}\geq\beta_1\sigma^2\label{con:secrecyrate}\\
     &\mathbf{w}^H\mathbf{B}_2\mathbf{w}\geq(2^{Q_s}-1)\sigma^2\\
     &\frac{\mathbf{w}^H\mathbf{B}_3\mathbf{w}}{\sigma^2}\leq\frac{1+\beta_1}{2^{Q_p}}-1\label{betaup}
\end{align}
has the same optimal solution to (P1),
where we define
\begin{align}
 &\mathbf{B}_1\triangleq\left(\begin{array}{cc}\mathbf{h}_{pp}\mathbf{h}_{pp}^H&\mathbf{0}\\\mathbf{0}& -\beta_1\mathbf{h}_{sp}\mathbf{h}_{sp}^H\end{array}\right),\\
     &\mathbf{B}_2\triangleq\left(\begin{array}{cc}-(2^{Q_s}-1)\mathbf{h}_{ps}\mathbf{h}_{ps}^H&\mathbf{0}\\\mathbf{0}& \mathbf{h}_{ss}\mathbf{h}_{ss}^H\end{array}\right),\\
     &\mathbf{B}_3\triangleq\left(\begin{array}{cc}\mathbf{h}_{ps}^H\mathbf{h}_{ps}&\mathbf{0}\\\mathbf{0} &\mathbf{0} \end{array}\right).
\end{align}

Let $g_1({\beta_1})$ denote the optimal value of (P1.1) with given $\beta_1$. It can thus be shown that (P1) achieves the same optimal value of the following problem:
\begin{align}
{\rm (P1.2)} ~~ ~\min_{\beta_1>0} g_1({\beta_1}).\nonumber
\end{align}

Therefore, the $\beta_1^*$ can be optimally solved by one dimension search over $\beta_1>0$. With any given $\beta_1$, $g_1(\beta_1)$ is obtained by solving (P1.1). Thus, in the following, we only need to focus on the solution for (P1.1).

By introducing $\mathbf{X}_1\triangleq\mathbf{w}\mathbf{w}^H$, (P1.1) can be equivalently rewritten as
\begin{align}
\min_{\mathbf{X}_1}~ &{\rm tr}\{\mathbf{X}_1\}\label{pro:SDP1}\\
{\rm s.t.} ~&{\rm tr}\{\mathbf{B}_1\mathbf{X}_1\}\geq \beta_1\sigma^2\\
&{\rm tr}\{\mathbf{B}_2\mathbf{X}_1\}\geq (2^{Q_s}-1)\sigma^2\\
&{\rm tr}\{\mathbf{B}_3\mathbf{X}_1\}\leq\sigma^2\left(\frac{1+\beta_1}{2^{Q_p}}-1\right)\\
&{\rm rank}\{\mathbf{X}_1\}=1\\
&\mathbf{X}_1\succeq\mathbf{0},\label{pro:SDP3}
\end{align}
where $\mathbf{X}_1\succeq\mathbf{0}$ means that $\mathbf{X}_1$ is a positive semidefinite (PSD) matrix and the above problem is a semidefinite programming (SDP) problem.

Note that the SDP problem in \eqref{pro:SDP1} is non-convex due to the the rank-one constraint. However, it can be solved by the semidefinite relaxation (SDR) technique \cite{SDR} as explained in the following. First, we drop the rank-one constraint to obtain the relaxed SDP problem as follows
\begin{align}
\min_{\mathbf{X}_1}~ &{\rm tr}\{\mathbf{X}_1\}\label{alg1}\\
{\rm s.t.} ~&{\rm tr}\{\mathbf{B}_1\mathbf{X}_1\}\geq \beta_1\sigma^2\\
&{\rm tr}\{\mathbf{B}_2\mathbf{X}_1\}\geq (2^{Q_s}-1)\sigma^2\\
&{\rm tr}\{\mathbf{B}_3\mathbf{X}_1\}\leq\sigma^2\left(\frac{1+\beta_1}{2^{Q_p}}-1\right)\\
&\mathbf{X}_1\succeq\mathbf{0}.\label{alg2}
\end{align}
The relaxed SDP problem in \eqref{alg1} can be solved conveniently by SDP solvers such as CVX \cite{CVX}. Due to the relaxation, $\mathbf{X}_1^*$ obtained by problem in \eqref{alg1} might not be rank-one in general, however, can be solved by the rank-one decomposition theorem \cite{HerDecom} given in the following lemma.

\textit{Lemma 1 [14, Theorem 2.3]:} Let $\mathbf{G}_i, i = 1,...,4$ be an $n \times n$ Hermitian matrix, and $\mathbf{X}$ be an $n \times n$ nonzero Hermitian PSD matrix. Suppose that $n\geq3$, if rank$(\mathbf{X})\geq3$, then one can find a nonzero vector $\mathbf{x}\in {\rm Range}(\mathbf{X})$ such that $\mathbf{G}_i \mathbf{xx}^H=\mathbf{G}_i\mathbf{X}, i = 1,...,4$. If rank$(\mathbf{X})=2$, then for any $\mathbf{y}\notin {\rm Range}(\mathbf{X})$, there exists $\mathbf{x}\in \mathbb{C}^{n\times1}$ in the linear subspace spanned by $\mathbf{y}$ and ${\rm Range}(\mathbf{X})$, such that $\mathbf{G}_i \mathbf{xx}^H=\mathbf{G}_i\mathbf{X}, i = 1,...,4$.

According to Lemma 1, we can recover the rank-one solution $\mathbf{w}^*$ that satisfies $\mathbf{w}^{H*}\mathbf{w}^{*}={\rm tr}(\mathbf{X}_1^*)$ and $\mathbf{w}^{H*}\mathbf{B}_{i}\mathbf{w}^{*}={\rm tr}(\mathbf{B}_{i}\mathbf{X}_1^*)$ for $i=1,2,3$ without loss of optimality of the SDR.

Finally, the proposed solution for (P1) is summarized in Algorithm 1.

\begin{algorithm}[tb]
\caption{Proposed Underlay Scheme Algorithm}
\begin{algorithmic}[1]
\STATE Set $\beta_1^{ub}$ and $M$ sufficiently large. Define $\beta_1^i=i\beta_1^{ub}/M$, $i=1,...,M$.
\FOR {Each $\beta_1^i$}
\STATE Set $\beta_1=\beta_1^i$.
\STATE Obtain an optimal $\mathbf{X}_1^*$ by solving problem \eqref{alg1}-\eqref{alg2} with given $\beta_1$.
\IF {rank($\mathbf{X}_1^*$)=1}
\STATE Decompose $\mathbf{X}_1^*=\mathbf{w}\mathbf{w}^H$.
\ELSE
\STATE Find $\mathbf{w}$ such that $\mathbf{w}^H\mathbf{w}={\rm tr}\{\mathbf{X}_1\}$, $\mathbf{w}^H\mathbf{B}_j\mathbf{w}={\rm tr}\{\mathbf{B}_j\mathbf{X}_1\}$, for $j=1,...,3$ according to Lemma 1.
\ENDIF
\ENDFOR
\STATE Select $\beta_1^i$ that achieves the minimal total power and the corresponding $\mathbf{w}^*$. Find $\mathbf{w}_1^*$ and $\mathbf{w}_2^*$ such that $[\mathbf{w}_1^{*H} ~\mathbf{w}_2^{*H}]^H=\mathbf{w}^*$.
\end{algorithmic}
\end{algorithm}
%

\section{Beamforming Design for Cooperative Scheme}

 In the following, we solve the (P2) for the cooperative scheme, where the PU and SU's beamforming vectors $\mathbf{w}_1$, $\mathbf{w}_2$ and the relay beamforming matrix at SU-Tx $\mathbf{F}$ should be jointly designed. However, (P2) is nonconvex due to the constraints in \eqref{con:secrecyrate1} and \eqref{con:rate}. Therefore we propose an iterative algorithm to solve (P2) efficiently based on the alternating optimization. In particular, we optimize $\mathbf{w}_1$ and $\mathbf{w}_2$ with fixed $\mathbf{F}$ and then solve $\mathbf{F}$ with fixed $\mathbf{w}_1$ and $\mathbf{w}_2$. The process is iterated until convergence. In the following subsections, we detail the derivations.
\subsection{Joint Users' Beamforming with given $\mathbf{F}$}
Given the relay beamforming matrix $\mathbf{F}$, the optimization problem (P2) can be reformulated as the follow QCQP
\begin{align}
{\rm (P2.1)} \nonumber&\\
\min_{\mathbf{w}_1,\mathbf{w}_2}\nonumber\label{P2.11}~ &\mathbf{w}_1^H(\mathbf{H}_0^H\mathbf{F}^H\mathbf{F}\mathbf{H}_0+\mathbf{I}_{N_p})\mathbf{w}_1+\mathbf{w}_2^H\mathbf{w}_2\\
{\rm s.t.} ~&\mathbf{w}_1^H\left[\frac{1}{\sigma^2}(\mathbf{h}_{pp}\mathbf{h}_{pp}^H-4^{Q_p
}\mathbf{H}_e^H\mathbf{H}_e)\right.\nonumber\\
&+\left.\frac{\mathbf{h}_{sp}^H\mathbf{F}\mathbf{H}_0\mathbf{H}_0^H\mathbf{F}^H\mathbf{h}_{sp}}{\mathbf{w}_{2}^H\mathbf{h}_{sp}\mathbf{h}_{sp}^H\mathbf{w}_{2}+\sigma^2(1+\mathbf{h}_{sp}^H\mathbf{F}\mathbf{F^H}\mathbf{h}_{sp})}\right]\mathbf{w}_1\geq 4^{Q_p}-1\\
&\frac{\mathbf{w}_2^H\mathbf{h}_{sp}\mathbf{h}_{sp}^H\mathbf{w}_2^H}{4^{Q_s}-1}-\mathbf{w}_1^H\mathbf{H}_0^H\mathbf{F}^H\mathbf{h}_{ss}\mathbf{h}_{ss}^H\mathbf{F}\mathbf{H}_0\mathbf{w}_1\nonumber\\
&\geq \sigma^2(1+\mathbf{h}_{ss}^H\mathbf{F}\mathbf{F}^H\mathbf{h}_{ss}). \label{con:remove}
\end{align}

Similar to the approach in the last section, we define $\mathbf{w}\triangleq[\mathbf{w}_1^H ~\mathbf{w}_2^H]^H$ to jointly design $\mathbf{w}_1$ and $\mathbf{w}_2$. It can be shown that there exists an interference temperature constraint $\beta_2$ that the following problem
\begin{align}
\min_{\mathbf{w}}~ &\mathbf{w}^H\mathbf{A}_1\mathbf{w}\label{P2.12}\\
{\rm s.t.} ~&\mathbf{w}^H\mathbf{B}_4\mathbf{w}\geq 4^{Q_p}-1\\
&\mathbf{w}^H\mathbf{B}_5\mathbf{w}\geq \sigma^2(1+\mathbf{h}_{ss}^H\mathbf{F}\mathbf{F}^H\mathbf{h}_{ss}) \label{con:remove}\\
&\mathbf{w}^H\mathbf{B}_6\mathbf{w}\leq\beta_2
\end{align}
has the same optimal value as problem in \eqref{P2.12}, where
\begin{align}
\mathbf{A}_1\triangleq&\left(\begin{array}{cc}\mathbf{H}_0^H\mathbf{F}^H\mathbf{F}\mathbf{H}_0+\mathbf{I}_{N_p}&\mathbf{0}\\\mathbf{0}& \mathbf{I}_{N_s}\end{array}\right),\\
\mathbf{B}_4\triangleq&\left(\begin{array}{cc}\mathbf{V}&\mathbf{0}\\
\mathbf{0}& \mathbf{0}\end{array}\right),\\
\mathbf{V}\triangleq&\frac{1}{\sigma^2}(\mathbf{h}_{pp}\mathbf{h}_{pp}^H-4^{Q_p
}\mathbf{H}_e^H\mathbf{H}_e)+\frac{\mathbf{h}_{sp}^H\mathbf{F}\mathbf{H}_0\mathbf{H}_0^H\mathbf{F}^H\mathbf{h}_{sp}}{\beta_2+\sigma^2(1+\mathbf{h}_{sp}^H\mathbf{F}\mathbf{F^H}\mathbf{h}_{sp})},\\
\mathbf{B}_5\triangleq&\left(\begin{array}{cc}-\mathbf{H}_0^H\mathbf{F}^H\mathbf{h}_{ss}\mathbf{h}_{ss}^H\mathbf{F}\mathbf{H}_0&\mathbf{0}\\  \mathbf{0}&\mathbf{h}_{ss}\mathbf{h}_{ss}^H/(4^{Q_s}-1)\end{array}\right),\\
\mathbf{B}_6\triangleq&\left(\begin{array}{cc}\mathbf{0}&\mathbf{0}\\\mathbf{0}&\mathbf{h}_{sp}\mathbf{h}_{sp}^H\end{array}\right).
\end{align}

By introducing $\mathbf{X}_2\triangleq\mathbf{w}\mathbf{w}^H$ and using the SDR technique, the problem can be equivalently rewritten as
%
\begin{align}
\min_{\mathbf{X}_2}~~&{\rm tr}\{\mathbf{A}_1 \mathbf{X}_2\}\label{alg3}\\
{\rm s.t.}~~ &{\rm tr}\{\mathbf{B}_4\mathbf{X}_2\}\geq 4^{Q_p}-1\\
&{\rm tr}\{\mathbf{B}_5\mathbf{X}_2\}\geq \sigma^2(1+\mathbf{h}_{ss}^H\mathbf{F}\mathbf{F}^H\mathbf{h}_{ss})\\
&{\rm tr}\{\mathbf{B}_6\mathbf{X}_2\}\leq\beta_2\\
&\mathbf{X}_2\succeq\mathbf{0}.\label{alg4}
\end{align}

After solving the above problem, we use Lemma 1 to recover the rank-one solution $\mathbf{w}^*$ with given $\beta_2$ for the case rank$(\mathbf{X}_2^*)>1$.
\subsection{Relay Beamforming Matrix $\mathbf{F}$ with Given $\mathbf{w}_1$ and $\mathbf{w}_2$}
Given the $\mathbf{w}_1$ and $\mathbf{w}_2$, the optimization problem for $\mathbf{F}$ is formulated as a QCQP:
\begin{align}
{\rm (P2.2)}~~\min_{\mathbf{F}} ~~&{\rm tr}\{\mathbf{F}\mathbf{H}_0\mathbf{w}_1\mathbf{w}_1^H \mathbf{H}_0^H\mathbf{F}^H+\sigma^2\mathbf{F}\mathbf{F}^H\}\nonumber\\
{\rm s.t.}~~ &\frac{1+\gamma_p}{1+\gamma_e}>4^{Q_p}\\
&\gamma_s>4^{Q_s}-1,
\end{align}
which can be further expressed as
\begin{align}
\min_{\mathbf{F}}~~&{\rm tr}\{\mathbf{F}^H\mathbf{F}(\mathbf{H}_0\mathbf{w}_1\mathbf{w}_1^H\mathbf{H}_0^H+\sigma^2\mathbf{I}_{N_s})\}\\
{\rm s.t.}~~&\frac{\mathbf{h}_{sp}^H\mathbf{F}\mathbf{H}_0\mathbf{w}_1\mathbf{w}_1^H\mathbf{H}_0^H\mathbf{F}^H\mathbf{h}_{sp}}{\mathbf{h}_{sp}^H\mathbf{w}_{2}\mathbf{w}_{2}^H\mathbf{h}_{sp}+\sigma^2(1+\mathbf{h}_{sp}^H\mathbf{F}\mathbf{F}^H\mathbf{h}_{sp})}
\geq\Omega,\\
&\frac{\mathbf{h}_{ss}^H\mathbf{w}_2\mathbf{w}_2^H\mathbf{h}_{ss}}{\sigma^2(1+\mathbf{h}_{ss}^H\mathbf{F}\mathbf{F}^H\mathbf{h}_{ss})+\mathbf{h}_{ss}^H\mathbf{F}\mathbf{H}_0\mathbf{w}_1\mathbf{w}_1^H\mathbf{H}_0^H\mathbf{F}^H\mathbf{h}_{ss}}\\
&\geq4^{Q_s}-1,
\end{align}
where
\begin{equation}
\Omega\triangleq4^{Q_p}-1+\frac{\mathbf{w}_1^H\left(4^{Q_p}\mathbf{H}_{e}^H\mathbf{H}_{e}-\mathbf{h}_{pp}\mathbf{h}_{pp}^H\right)\mathbf{w}_1}{\sigma^2}.
\end{equation}
Using ${\rm tr}(\mathbf{A}^H\mathbf{BAC})={\rm vec}(\mathbf{A})^H(\mathbf{C}^H\otimes \mathbf{B}){\rm vec}(\mathbf{A})$, we have
\begin{align}
\min_{\mathbf{f}}~~&\mathbf{f}^H \mathbf{A}_2\mathbf{f}\label{min:F}\\
{\rm s.t.}~~ &\mathbf{f}^H\mathbf{B}_7\mathbf{f}\geq \Omega(\mathbf{h}_{sp}^H\mathbf{w}_{2}\mathbf{w}_{2}^H\mathbf{h}_{sp}+\sigma^2)\label{con:F}\\
&\mathbf{f}^H\mathbf{B}_8\mathbf{f}\leq \frac{\mathbf{h}_{ss}^H\mathbf{w}_{2}\mathbf{w}_{2}^H\mathbf{h}_{ss}}{4^{Q_s}-1}-\sigma^2,
\end{align}
where
\begin{align}
\mathbf{f}&\triangleq{\rm vec}\mathbf {(F)}, \\ \mathbf{A}_2&\triangleq(\mathbf{H}_0\mathbf{w}_1\mathbf{w}_1^H\mathbf{H}_0^H+\sigma^2\mathbf{I}_{N_s})^H\otimes\mathbf{I}_{N_s},\\
\mathbf{B}_7&\triangleq(\mathbf{H}_0\mathbf{w}_1\mathbf{w}_1^H\mathbf{H}_0^H-\sigma^2\Omega\mathbf{I}_{N_s})^H\otimes(\mathbf{h}_{sp}\mathbf{h}_{sp}^H), \\ \mathbf{B}_8&\triangleq(\mathbf{H}_0\mathbf{w}_1\mathbf{w}_1^H\mathbf{H}_0^H+\sigma^2\mathbf{I}_{N_s})^H\otimes(\mathbf{h}_{ss}\mathbf{h}_{ss}^H).
\end{align}%
\begin{algorithm}[tb]
\caption{Proposed Cooperative Scheme Algorithm}
\begin{algorithmic}[1]
\STATE Initialize $\mathbf{F}{(0)}$; set $t=0$.
\STATE Set $\beta_2^{ub}$ and $M$ sufficiently large. Define $\beta_2^i=i\beta_2^{ub}/M$, $i=1,...,M$.
\REPEAT
\STATE $t=t+1$.
\FOR {each $\beta_2^i$}
\STATE Set $\beta_2=\beta_2^i$.
\STATE Obtain an optimal $\mathbf{X}_2^*$ by solving problem \eqref{alg1}-\eqref{alg2} with given $\beta_2$ and $\mathbf{F}{(t-1)}$.
\IF {rank($\mathbf{X}_2^*$)=1}
\STATE Decompose $\mathbf{X}_2^*=\mathbf{w}\mathbf{w}^H$.
\ELSE
\STATE Find $\mathbf{w}$ such that $\mathbf{w}^H\mathbf{w}={\rm tr}\{\mathbf{X}_2\}$, $\mathbf{w}^H\mathbf{B}_j\mathbf{w}={\rm tr}\{\mathbf{B}_j\mathbf{X}_2\}$, for $j=4,5,6$ and according to Lemma 1.
\ENDIF
\ENDFOR
\STATE Select $\beta_2^i$ that achieves the minimal total power and its corresponding $\mathbf{w}^*$. Find $\mathbf{w}_1(t)$ and $\mathbf{w}_2{(t)}$ such that $[\mathbf{w}_1^{H}{(t)} ~\mathbf{w}_2^{H}{(t)}]^H=\mathbf{w}^*$.
\STATE Obtain an optimal $\mathbf{X}_3^*$ by solving problem in \eqref{min1:F}-\eqref{con2:f}.
\IF {rank($\mathbf{X}_3^*$)=1}
\STATE Decompose $\mathbf{X}_3^*=\mathbf{f}\mathbf{f}^H$.
\ELSE
\STATE Find $\mathbf{f}$ such that $\mathbf{f}^H\mathbf{A}_2\mathbf{f}={\rm tr}\{\mathbf{A}_2\mathbf{X}_3\}$, $\mathbf{f}^H\mathbf{B}_j\mathbf{f}={\rm tr}\{\mathbf{B}_j\mathbf{X}_3\}$, for $j=7,8$ according to Lemma 1.
\ENDIF
\STATE Find $\mathbf{F}(t)$ such that $\mathbf{f}={\rm vec}(\mathbf{F}(t))$.
\UNTIL the total power converges.
\end{algorithmic}
\end{algorithm}
By introducing $\mathbf{X}_3=\mathbf{f}\mathbf{f}^H$ and using the SDR technique, the above problem can be reformulated as
\begin{align}
\min_{\mathbf{X}_3}~~&{\rm tr}\{ \mathbf{A}_2\mathbf{X}_3\}\label{min1:F}\\
{\rm s.t.}~~ &{\rm tr}\{ \mathbf{B}_7\mathbf{X}_3\}\geq \Omega(\mathbf{h}_{sp}^H\mathbf{w}_{2}\mathbf{w}_{2}^H\mathbf{h}_{sp}+\sigma^2)\label{con1:F}\\
&{\rm tr}\{ \mathbf{B}_8\mathbf{X}_3\}\leq \frac{\mathbf{h}_{ss}^H\mathbf{w}_{2}\mathbf{w}_{2}^H\mathbf{h}_{ss}}{4^{Q_s}-1}-\sigma^2.\label{con2:f}
\end{align}
For the case that $\mathbf{X}_3^*$ is not rank-one, the optimal rank-one solution of the problem in \eqref{min1:F} can be recovered by Lemma 1 without loss of optimality.

Finally, the above proposed solution for (P2) is summarized in Algorithm 2.

%

\section{Numerical Results}

In this section, we carry out simulations to evaluate the performance of our cooperative scheme and underlay scheme. We assume that the small-scale fading in each link follows independent Rayleigh distribution and the large-scale fading on each link is modeled by the path loss model as $\alpha_{i,j}=c\cdot d_{i,j}^{-n}$, where $d_{i,j}$ denotes the distance between $i\in\{$PU-Tx, SU-Tx$\}$ and $j\in\{$PU-Rx, SU-Rx$\}$, $c$ is the attenuation constant set to be 1 and $n$ is the path loss exponent set to be 3. Throughout this paper, we consider a system topology where the PU-Tx, PU-Rx, SU-Tx and SU-Rx are placed at $(-0.5,0)$, $(0.5,0)$, $(0,0)$ and $(0,0.5)$, respectively.  The noise power $\sigma^2$ is set to be 1.

\begin{figure}[t]
\begin{centering}
\includegraphics[scale=0.6]{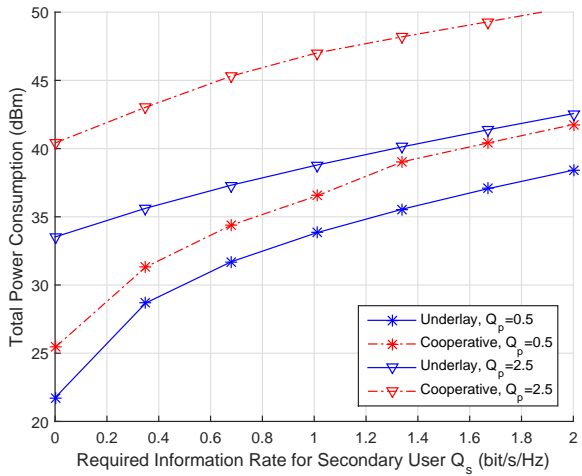}
\vspace{-0.1cm}
 \caption{The total power consumption versus secondary user's required information rate $Q_s$ with $Q_p=0.5$bit/s/Hz, $N_p=4$ and $N_s=2$. }\label{fig1}
\end{centering}
\vspace{-0.3cm}
\end{figure}

In Fig. \ref{fig1}, the total power consumption is plotted as a function of $Q_s$, where $N_p=4$, $N_s=2$. Two different primary secrecy rate requirement with $Q_p=0.5$bit/s/Hz and $Q_p=2.5$bit/s/Hz are simulated for each scenario. It can be first observed that with the increase of $Q_s$, the total consumed power of the system for all schemes increases to satisfy the SU's requirement. In addition, it can be shown that for both $Q_p=0.5$bit/s/Hz and $Q_p=2.5$bit/s/Hz scenarios, the underlay scheme outperforms the cooperative scheme. This indicates that under the considered scenarios, the cooperation with an untrusted SU may be unsafe since SU is capable to eavesdrop more confidential message from PU-Tx. Moreover, one can also observe that with the increase of $Q_p$ or $Q_s$, the gap between underlay scheme and cooperative scheme becomes larger.
    \begin{figure}[t]
\begin{centering}
    \includegraphics[scale=0.6]{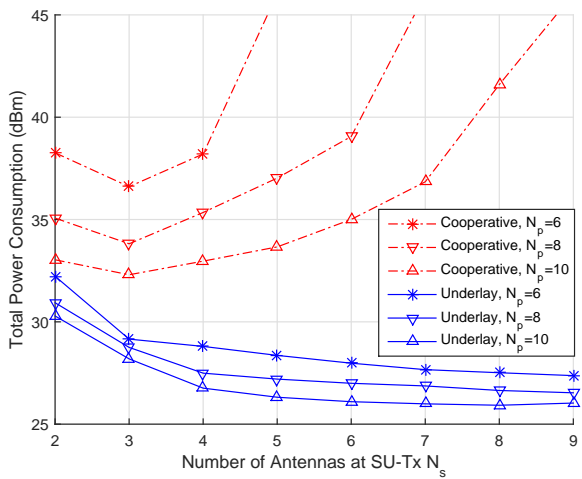}
\vspace{-0.1cm}
 \caption{The total power versus the number of the SU's antennas with $Q_p=2$bit/s/Hz and $Q_s=2$bit/s/Hz.}\label{fig2}
\end{centering}
\vspace{-0.3cm}
\end{figure}

Fig. \ref{fig2} demonstrates the total power consumption versus the number of antennas of the SU-Tx $N_s$ for both schemes, where we fix $Q_p=2$bit/s/Hz and $Q_s=2$bit/s/Hz with different number of the PU-Tx's antennas $N_p$. It is first observed that given $N_s$, both schemes with larger $N_p$ performs better. An interesting observation is that with the increase of $N_s$, the total power consumption for the cooperative scheme first decreases to the optimal point (when $N_s=3$) and then increases sharply when $N_s$ becomes closer to $N_p$. The reason is that, with greater spatially diversity gain provided by more antennas, the SU can consume less power for both relaying the PU's information and transmitting its own message. However, when $N_s$ becomes greater than $3$, the SU becomes much more capable of eavesdropping PU's confidential message. Thus, the PU and SU have to consume much more power to guarantee the PU's required secrecy rate. It is also shown that the underlay scheme has better performance as $N_s$ become larger.

\section{Conclusions}
In this paper, we studied a CRN in the presence of an untrusted SU with the aim of minimizing the total transmit power while satisfying the requirements of both PU and SU. The underlay scheme and the cooperative scheme were studied and we designed the corresponding joint beamforming for both schemes using suitable optimization tools. The simulation results showed that the underlay scheme outperforms the cooperative scheme, especially with high rate requirements and a large number of antennas at SU-Tx.

\bibliographystyle{IEEEtran}
\bibliography{IEEEabrv,Untrusted_Cogntive}

\end{document}